\documentclass[cits]{PoS}
\usepackage[sort&compress]{natbib}
\bibpunct{[}{]}{,}{n}{}{}

\newcommand{\ev}{\, {\rm eV}}

\newcommand{\comment}[1]{}
\newcommand{\lr}[1]{ \left( #1 \right) }
\newcommand{\lrs}[1]{ \left[ #1 \right] }

\newcommand{\tr}{ {\rm Tr} \, }
\newcommand{\re}{ {\rm Re} \, }

\renewcommand{\det}[1]{ {\rm det} \left( #1 \right) }

\newcommand{\expa}[1]{ \exp{\left( #1 \right)} }

\title{Monte-Carlo study of quasiparticle dispersion relation in monolayer graphene}

\ShortTitle{Quasiparticle dispersion relation in graphene}

\author{\speaker{P. V. Buividovich}\thanks{This work was supported by the S. Kowalewskaja award from the Alexander von Humboldt Foundation.}\\
        Institute for Theoretical Physics, Regensburg University, Universit\"{a}tsstrasse 31, D-93053 Regensburg, Germany\\
        E-mail: \email{pavel.buividovich@physik.uni-regensburg.de}}

\abstract{The density of electronic one-particle states in monolayer graphene is studied by performing the Hybrid Monte-Carlo simulations of the tight-binding model for electrons on the $\pi$ orbitals of carbon atoms which make up the graphene lattice. Density of states is approximated as a derivative of the number of particles over the chemical potential at sufficiently small temperature. Simulations are performed in the partially quenched approximation, in which virtual particles and holes have zero chemical potential. It is found that the Van Hove singularity becomes much sharper than in the free tight-binding model. Simulation results also suggest that the Fermi velocity increases with interaction strength up to the transition to the phase with spontaneously broken chiral symmetry.}

\FullConference{Xth Quark Confinement and the Hadron Spectrum,\\
		October 8-12, 2012\\
		TUM Campus Garching, Munich, Germany}

\begin{document}

\section{Introduction}

\begin{figure}
  \centering
  \includegraphics[width=14cm]{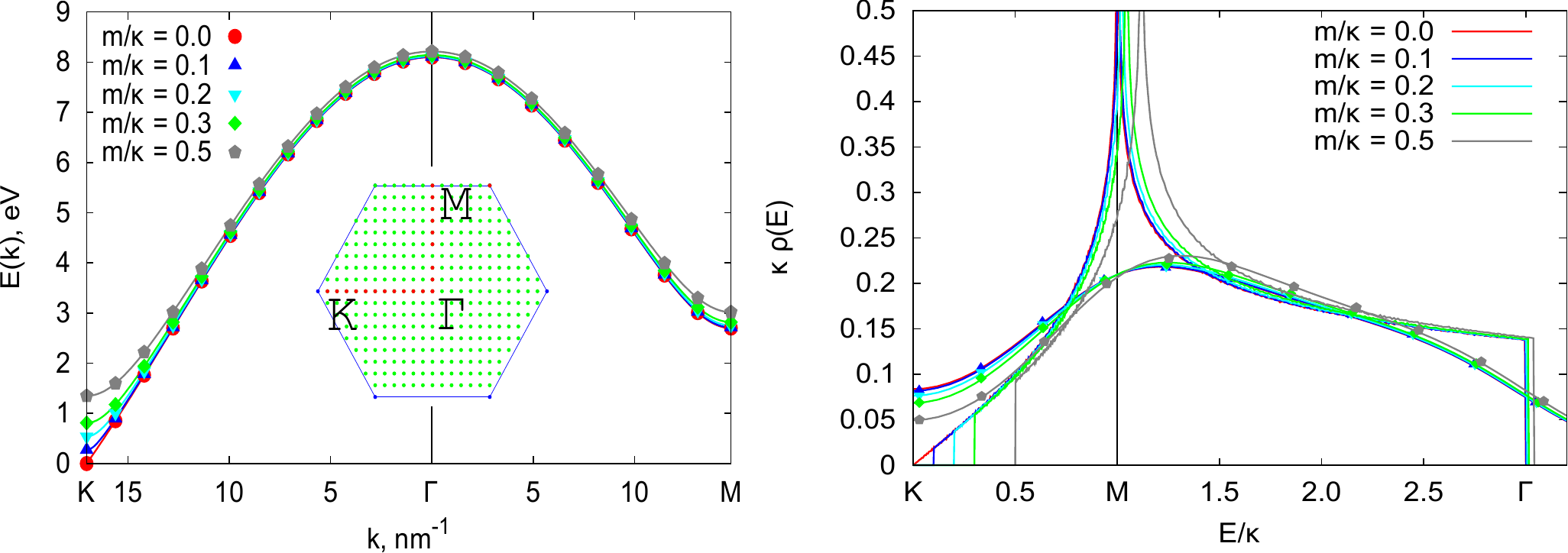}\\
  \caption{A cut of the dispersion relation (left) and the density of states (right) for the non-interacting tight-binding model with different values of the staggered potential $m$. The inset on the left plot shows the filling of the hexagonal Brillouin zone of graphene with discrete lattice momenta on the $18 \times 18$ lattice with periodic boundary conditions and the positions of the Dirac point ($K$), Van Hove singularity ($M$) and the edge of the valence band ($\Gamma$). These points are also marked on the right plot (for $m = 0$). Solid lines with points on them illustrate the convolution (1.2) of the density of states on the $18 \times 18$ lattice with the derivative of the Fermi factor at $T/\kappa = 0.28$. Lines correspond to the results of exact calculation and points - to the numerical results obtained with the discretized fermionic operator (see below).}
  \label{fig:dispersion}
\end{figure}

 Electron transport properties of graphene are of great importance both for the industrial applications of this novel material as well as for our understanding of the physics of strongly correlated electrons. It turns out that at low energies electronic excitations in graphene effectively behave as free massless Dirac fermions \cite{Novoselov:09:1}. However, they propagate with the Fermi velocity $v_F = c/300$, which is much less than the speed of light $c$. There is also a natural way to make these Dirac fermions massive: one should break the symmetry between the two simple rhombic sublattices of hexagonal graphene lattice by introducing the so-called ``staggered potential'', so that the potential is lifted by some value $m$ on the sites of one sublattice and lowered by $m$ on the sites of other sublattice. This ``staggered potential'' $m$ then plays the role of the Dirac mass. At higher energies of order of the hopping energy $\kappa \approx 2.8 \ev$ the quasiparticle dispersion relation becomes nonlinear. At $E = \pm \sqrt{\kappa^2 + m^2}$, it has a saddle point, which results in the logarithmic divergence in the density of states. This Van Hove singularity leads to the so-called Lifshitz transition as the Fermi energy of the system approaches the saddle point. The total width of the valence band is $2 \sqrt{9 \kappa^2 + m^2}$. The dispersion relation and the density of states at different values of the staggered potential $m$ are illustrated on Fig. \ref{fig:dispersion}.

 Charge carriers in graphene are also subject to electromagnetic interactions. The treatment of these interactions can be significantly simplified due to the smallness of the Fermi velocity. First, retardation effects can be neglected and, second, the Lorentz force between the electrons also is suppressed by a factor $v_F/c$ as compared to the electric force. Thus it is sufficient to consider only the instantaneous Coulomb interaction between Dirac quasiparticles.

 An important question is how the parameters of the free tight-binding model such as the widths of the valence band, the Fermi velocity and the mass gap $m$ change when one takes into account the interactions between quasiparticles. In order to pose such a question, one has to adapt the Fermi-Landau liquid model and assume that despite the interactions, in some sense quasiparticles can be still described as free fermions with some modified dispersion relation. Strictly speaking, graphene is an example of the so-called marginal Fermi liquid \cite{Varma:89:1}, for which the Fermi-Landau liquid approximation becomes inapplicable in the vicinity of the Fermi point (Fermi point is a degenerate Fermi surface with zero radius) . However, as discussed in \cite{DasSarma:07:1}, even a small doping brings monolayer graphene back into the Fermi-Landau liquid regime. One can expect that the introduction of the ``staggered potential'', which eliminates the Fermi point, should have the same effect.

 In this paper the density of one-particle states in monolayer graphene is studied numerically. It is found that the Lifshitz phase transition, associated with the Van Hove singularity around the saddle point in the dispersion relation of the tight-binding model, becomes much more pronounced in the interacting theory. Such behavior is in agreement with the analytical calculations based on renormalization-group arguments \cite{Vozmediano:97:1}. There are also indications that the Fermi velocity increases with the interaction strength up to the transition to the phase with spontaneously broken chiral symmetry, again in agreement with renormalization-group arguments \cite{Shankar:94:1}. On the other hand, it seems that the width of the artificially induced (with the help of the staggered potential) mass gap remains practically constant. However, measurements with a better energy resolution are required to clarify the situation completely.

 In analytical calculations, it is most convenient to study the quasiparticle dispersion relation and the corresponding density of states by analyzing the poles of the fermionic Green functions. In numerical Monte-Carlo simulations, however, fermionic Green functions can only be calculated for a finite number of discrete values of imaginary (Euclidean) time. Extraction of the real-time Green functions from such Euclidean field correlators is an ill-defined numerical problem, which has no unique solution. The most commonly used method for extracting real-time quantities from Euclidean correlators is the Maximal Entropy Method (MEM) \cite{Gubernatis:96:1}. However, this method introduces significant systematic uncertainties, in particular, due to ambiguity of the choice of the ``model function'' which incorporates our prior knowledge about the expected form of the corresponding spectral function. Typically, MEM tends also to smear the singularities (such as thresholds or Van Hove singularities) of the spectral functions at the scales of order of temperature. An attempt to extract the AC conductivity of graphene from numerically calculated current-current correlators was already reported by the author in \cite{Buividovich:12:1}. It was found that MEM indeed was not able to reproduce the AC conductivity with precision which would be sufficient for quantitative analysis.

 It should be noticed, however, that in a strict sense the density of states which can be formally extracted from the fermionic Green function does not correspond to the density of any eigenstates of the interacting Hamiltonian. The reason is that the notion of quasiparticle is not strictly defined in this case. The interpretation of the spectral function of interacting fermionic gas in terms of the density of quasiparticle states only makes sense in the framework of the Landau-Fermi liquid model. However, in this framework one can also think of many other possible definitions of the density of states, which are only constrained by the requirement to correctly reproduce the density of states for a free Fermi gas with an arbitrary dispersion relation for fermions.

 In this work, the density of states is defined as the derivative of the number of particles over chemical potential. For a gas of free fermions with the density of one-particle states $\rho\lr{E}$, the particle density $n\lr{\mu}$ and its derivative $\frac{d n\lr{\mu}}{d \mu}$ can be written as:
\begin{eqnarray}
\label{n_vs_mu}
 n\lr{\mu} = \int dE \, \rho\lr{E} \, \frac{1}{1 + \expa{\frac{E - \mu}{k T}}} \\
\label{dndmu_vs_mu}
 \frac{d n\lr{\mu}}{d \mu} = \frac{1}{k T}
 \int dE \, \rho\lr{E} \, \frac{1}{4 \cosh^2\lr{\frac{E - \mu}{2 k T}}}
\end{eqnarray}
In the limit of zero temperature, the Fermi factor $\frac{1}{1 + \expa{\frac{E - \mu}{k T}}}$ becomes the Heaviside step function $\theta\lr{\mu - E}$, and its derivative with respect to the chemical potential becomes the $\delta$-function $\delta\lr{\mu - E}$. Thus in the limit of zero temperature the derivative $dn\lr{\mu}/d\mu$ is exactly the density of one-particle states. At finite temperature the $\delta$-function is smeared over a finite energy range with the width of order of temperature and becomes an exponentially decaying function. A direct estimate of the distribution width yields:
\begin{eqnarray}
\label{fermi_distribution_width}
 \frac{1}{k T}
 \int dE \, \frac{\lr{E - \mu}^2}{4 \cosh^2\lr{\frac{E - \mu}{2 k T}}} = \frac{\pi^2 \lr{k T}^2}{3}
\end{eqnarray}
Thus by performing simulations at sufficiently small temperatures and by measuring the derivative $\frac{d n\lr{\mu}}{d \mu}$, one can obtain a reasonably accurate estimate of the density of one-particle states. The resulting function is smeared as compared to the zero-temperature limit, similarly to the results which can be obtained by using MEM. On the other hand, such an estimate is not based on any ``model function'' and is thus free from possible bias in the measurements caused by the particular choice of this function.

 In practice, a direct numerical implementation of such measurements is a formidably difficult task due to the so-called ``sign problem'' of the Monte-Carlo simulations at finite chemical potential \cite{Troyer:05:1}. At zero chemical potential (that is, for graphene at half-filling) the sign problem is absent due to the symmetry between particles and holes \cite{Lahde:09:2, Buividovich:12:1, Rebbi:12:1}. For this reason, in this work the quantity $\frac{d n\lr{\mu}}{d \mu}$ is calculated in the partially quenched approximation, in which virtual particles are not influenced by finite chemical potential. Such approximation can be introduced as follows: consider the tight-binding model with $1 + N_f$ independent fermion flavors and assume that chemical potential is nonzero only for $N_f$ fermions. For the time being it is convenient to assume that the staggered potential $m$ is equal to zero. Transforming the partition function of such a system into the path integral representation along the lines of \cite{Buividovich:12:1, Lahde:09:2, Rebbi:12:1}, it is straightforward to obtain the following result:
\begin{eqnarray}
\label{partition_function_path_int}
 \mathcal{Z}\lr{\mu, T, N_f} = \int\mathcal{D}\phi \,
 |\det{M}|^2 \det{M + \mu}^{N_f} \overline{\det{M - \mu}}^{N_f} e^{-S\lrs{\phi }} ,
\end{eqnarray}
where $\phi \equiv \phi\lr{X, \tau}$ is the Hubbard-Stratonovich \cite{Rebbi:12:1} or the electrostatic potential \cite{Buividovich:12:1, Lahde:09:2} field with the action $S\lrs{\phi }$, $\tau \in \lrs{0, \lr{kT}^{-1}}$ is the Euclidean time, $X$ variable labels the sites of the graphene hexagonal lattice, $M = \partial_{\tau} - h_0 + i \phi\lr{X, \tau}$ is the fermionic hopping operator and $h_0$ is the one-particle Hamiltonian of the tight-binding model. Consider now the linear response of the system to adding more flavours of fermions at finite chemical potential, that is, the derivative $\frac{\partial \log \mathcal{Z}\lr{\mu, T, N_f}}{\partial N_f}|_{N_f \rightarrow 0}$. This yields the partially quenched partition function
\begin{eqnarray}
\label{part_quenched_part_func}
 \mathcal{Z}_q\lr{\mu, T} = \mathcal{Z}^{-1}\lr{\mu=0, T, N_f = 0} \, \int\mathcal{D}\phi
 |\det{M}|^2 \, 2 \re\tr\log\lr{M + \mu} \, e^{-S\lrs{\phi }}  .
\end{eqnarray}
This expression was derived by taking into account the invariance of the path integral under the reflection $\phi \rightarrow -\phi$ and the anti-unitary symmetry of the operator $M$ at zero chemical potential: $C M C = -M^{\dag}$ with $C = C\lr{X, \tau_1; Y, \tau_2} = \pm \delta_{XY} \delta\lr{\tau_1, \tau_2}$, where the plus and the minus signs are taken for even and odd sites of the graphene hexagonal lattice, respectively. Clearly, the square of $C$ is just the identity operator. At low energies, when quasiparticles in graphene can be described as Dirac fermions, $C$ plays the role of the $\gamma_5$ Dirac matrix.

 Finally, taking the derivatives of the quenched partition function over the chemical potential, one finds the partially quenched approximation for $n\lr{\mu}$ and $d n\lr{\mu}/d\mu$:
\begin{eqnarray}
\label{n_vs_mu_part_quench}
 n_q\lr{\mu} = \mathcal{Z}^{-1}\lr{\mu=0, T, N_f = 0} \, \int\mathcal{D}\phi
 |\det{M}|^2 \, \frac{2 k T}{V} \, \re\tr\lr{M + \mu}^{-1} \, e^{-S\lrs{\phi }}
 \\
\label{dndmu_vs_mu_part_quench}
 \frac{d n_q\lr{\mu}}{d \mu} = \mathcal{Z}^{-1}\lr{\mu=0, T, N_f = 0} \, \int\mathcal{D}\phi
 |\det{M}|^2 \, \frac{2 k T}{V} \, \re\tr\lr{M + \mu}^{-2} \, e^{-S\lrs{\phi }}  ,
\end{eqnarray}
where $V$ is the number of lattice sites in the system. These expressions differ from the expression for the full, unquenched functions $n\lr{\mu}$ and $dn\lr{\mu}/d\mu$ by the absence of chemical potential in the factor $|\det{M}|^2$. However, since the density of one-particle states is not a rigorously defined quantity in the interacting theory, there is no reason to believe that $dn\lr{\mu}/d\mu|_{\mu = E}$ is a better estimate of $\rho\lr{E}$ then $dn_q\lr{\mu}/d\mu|_{\mu = E}$. Indeed, for a free fermion gas both definitions are completely equivalent.

 The observables (\ref{n_vs_mu_part_quench}) and (\ref{dndmu_vs_mu_part_quench}) are now well-suited for numerical simulations and can be calculated by averaging the quantities $\lr{M + \mu}^{-1}$ and $\lr{M + \mu}^{-2}$ over the equilibrium ensemble of the fields $\phi\lr{X, \tau}$ with the weight $|\det{M}|^2 \, e^{-S\lrs{\phi}}$ \cite{Lahde:09:2, Rebbi:12:1, Buividovich:12:1}. To this end Hybrid Monte-Carlo simulations of the tight-binding model of graphene on the lattice with toric topology which consisted of $18 \times 18$ hexagonal cells were performed. Temporal size of the lattice is $L_{\tau} = 18$ with lattice spacing $\kappa {\Delta \tau} = 0.2$. Coulomb interactions are modelled by coupling the tight-binding model to the $U\lr{1}$ noncompact gauge field in $3 + 1$ dimensions, as in \cite{Buividovich:12:1}. The corresponding coupling constant is controlled by the dielectric permittivity $\epsilon$ of substrate on which the graphene monolayer is placed: $\alpha = 2 \alpha_0/\lr{v_F \, \lr{\epsilon + 1}}$, where $\alpha_0 = e^2/4\pi = 1/137$ \cite{Buividovich:12:1}. Lattice size in the direction perpendicular to the graphene plane is $L_z = 18$. Simulation setup and the discretization of the fermionic operator $M$ are the same as in \cite{Buividovich:12:1}. Despite the fact that the estimate $\frac{d n_q\lr{\mu}}{d \mu}$ of the density of states in (\ref{dndmu_vs_mu_part_quench}) can be expressed in terms of the partially quenched partition function (\ref{part_quenched_part_func}) only at zero staggered potential $m$, the simulations have to be performed at some nonzero $m$ (in this work $m/\kappa = 0.1$ and $m/\kappa = 0.5$) in order to ensure the invertibility of the fermionic operator $M$. The derivative $\frac{d n_q\lr{\mu}}{d \mu}$ is found by first calculating $n_q\lr{\mu}$ according to (\ref{n_vs_mu_part_quench}) for equidistant values of $\mu$ separated by $\Delta \mu = 0.2 \, \kappa$ and then numerically differentiating this function using the symmetric difference
\begin{eqnarray}
\label{symmetric_difference}
 \frac{d n_q\lr{\mu}}{d \mu} \approx \frac{n_q\lr{\mu + {\Delta \mu}} - n_q\lr{\mu - {\Delta \mu}}}{2 {\Delta \mu}} .
\end{eqnarray}

\begin{figure}[htb]
  \centering
  \includegraphics[type=pdf,ext=.pdf,read=.pdf,width=7cm,angle=-90]{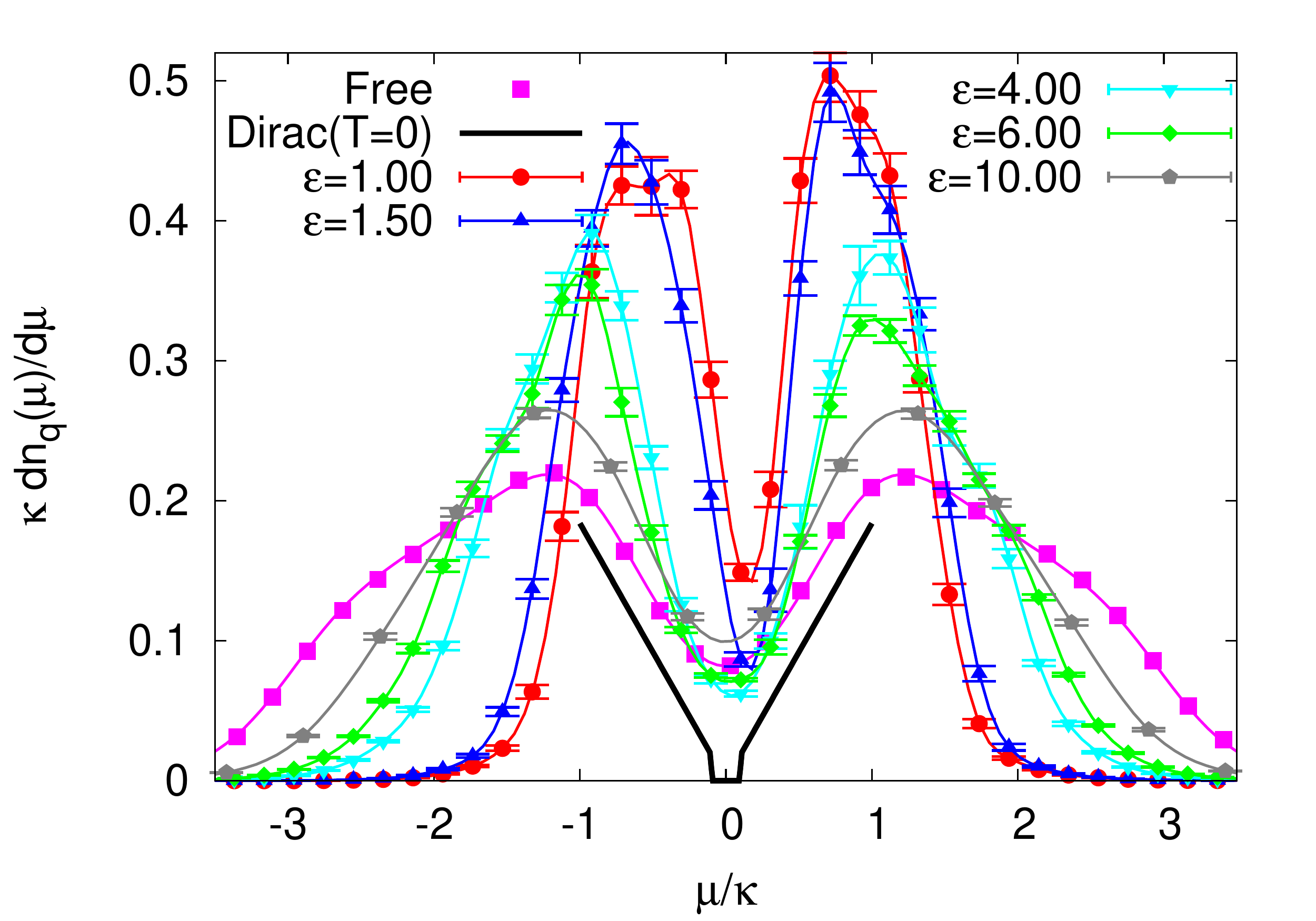}\\
  \includegraphics[type=pdf,ext=.pdf,read=.pdf,width=7cm,angle=-90]{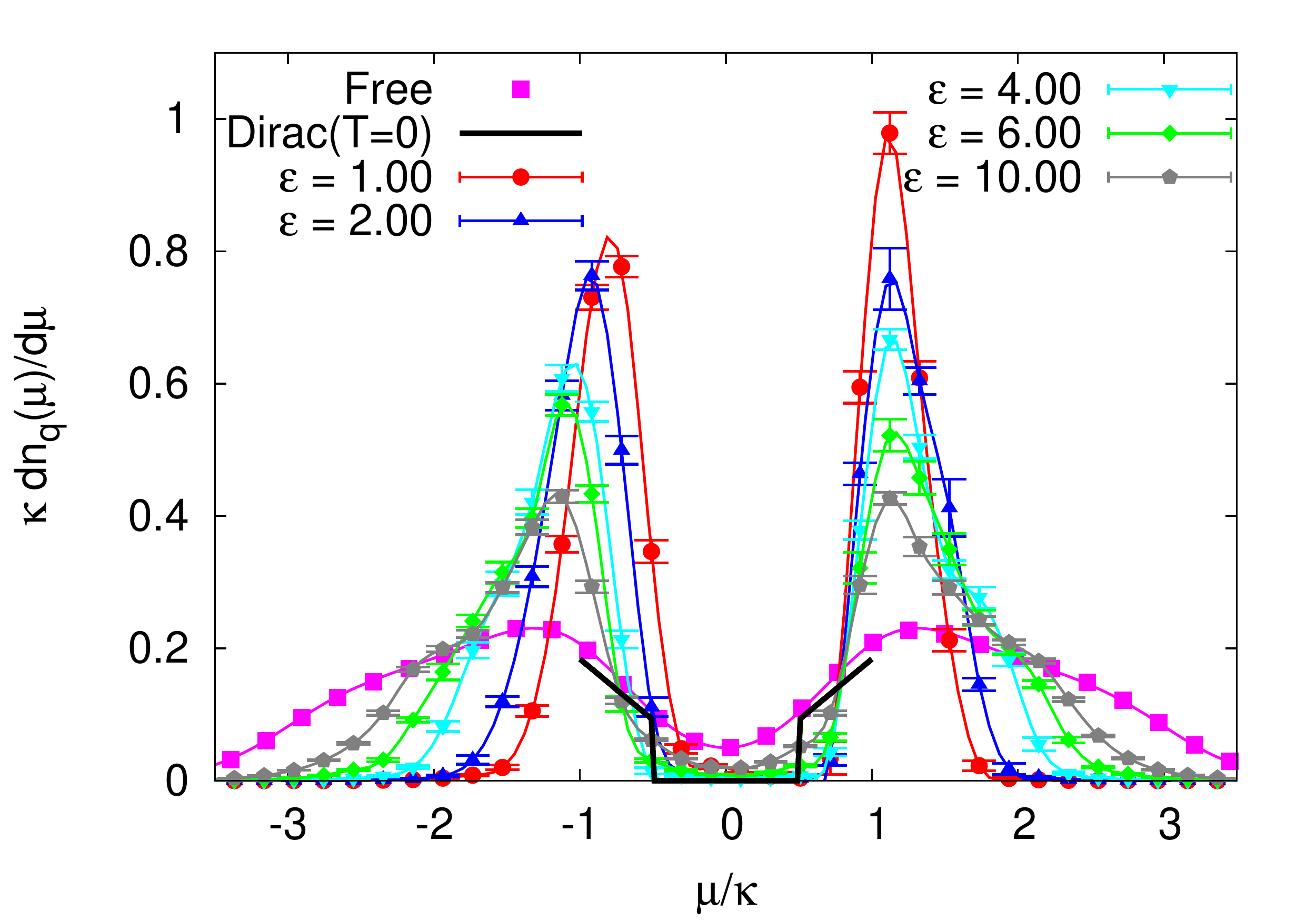}\\
  \caption{Estimate (1.7) of the density of states for the $18 \times 18$ lattice with $\kappa {\Delta \tau} = 0.2$ ($T/\kappa = 0.28$) and different values of the substrate dielectric permittivity at $m/\kappa = 0.1$ (above) and $m/\kappa = 0.5$ (below). Solid lines connecting the data points are cubic splines which are plotted to guide the eye.}
  \label{fig:dndmu_k0.2000_s18_t18}
\end{figure}

 The resulting dependence of $d n_q\lr{\mu}/d \mu$ on $\mu$ is illustrated on Fig. \ref{fig:dndmu_k0.2000_s18_t18} for $m/\kappa = 0.1$ and $m/\kappa = 0.5$. For comparison, the function $d n_q\lr{\mu}/d \mu$ for the free theory and the density of states $\rho_{Dirac}\lr{E}$ for free Dirac fermions (which corresponds to $d n_q\lr{\mu}/d \mu$ in the limit of zero temperature) are also plotted:
\begin{eqnarray}
\label{dos_dirac}
 \rho_{Dirac}\lr{E} = \frac{3 \sqrt{3} a^2}{8 \pi v_F^2} \, |E| , \quad |E| \ge m, \quad \quad \rho_{Dirac}\lr{E} = 0, \quad |E| < m  .
\end{eqnarray}
This expression takes into account that on the hexagonal lattice with periodic boundary conditions the number of states in the element $d^2 k$ of the momentum space is given by $dN/V = \frac{3 \sqrt{3} a^2}{4 \lr{2 \pi}^2} \, d^2 k$ \cite{Buividovich:12:1}.

 First of all one can note the sharp rise of the peaks at $E \approx \pm \kappa$, which are associated with the Van Hove singularities in the free theory. This is an indication that the Lifshitz phase transition (associated with the crossing of the Van Hove singularity by the Fermi level) might become much sharper in the interacting theory. Indeed, such phase transitions are known to be unstable with respect to even small interaction between particles. Application of renormalization-group techniques to quasiparticles in the vicinity of the Van-Hove singularities also show that as the interactions are switched on, the density of states tends to increase \cite{Vozmediano:97:1}. In other words, the quasiparticle dispersion relation becomes more flat in the vicinity of the $M$ point. This sharpening of the Van Hove singularity can be clearly seen for both values of $m$ starting from the smallest nonzero coupling constant (which corresponds to substrate dielectric permittivity $\epsilon = 10.0$). It is also interesting to note a small asymmetry between the Van Hove singularities at $E = -\kappa$ and at $E = \kappa$ and a slight shift of the position of the minimum of the density of states from $E = 0$ to positive values of $E$.

\begin{figure}[htb]
  \centering
  \includegraphics[type=pdf,ext=.pdf,read=.pdf,width=6cm,angle=-90]{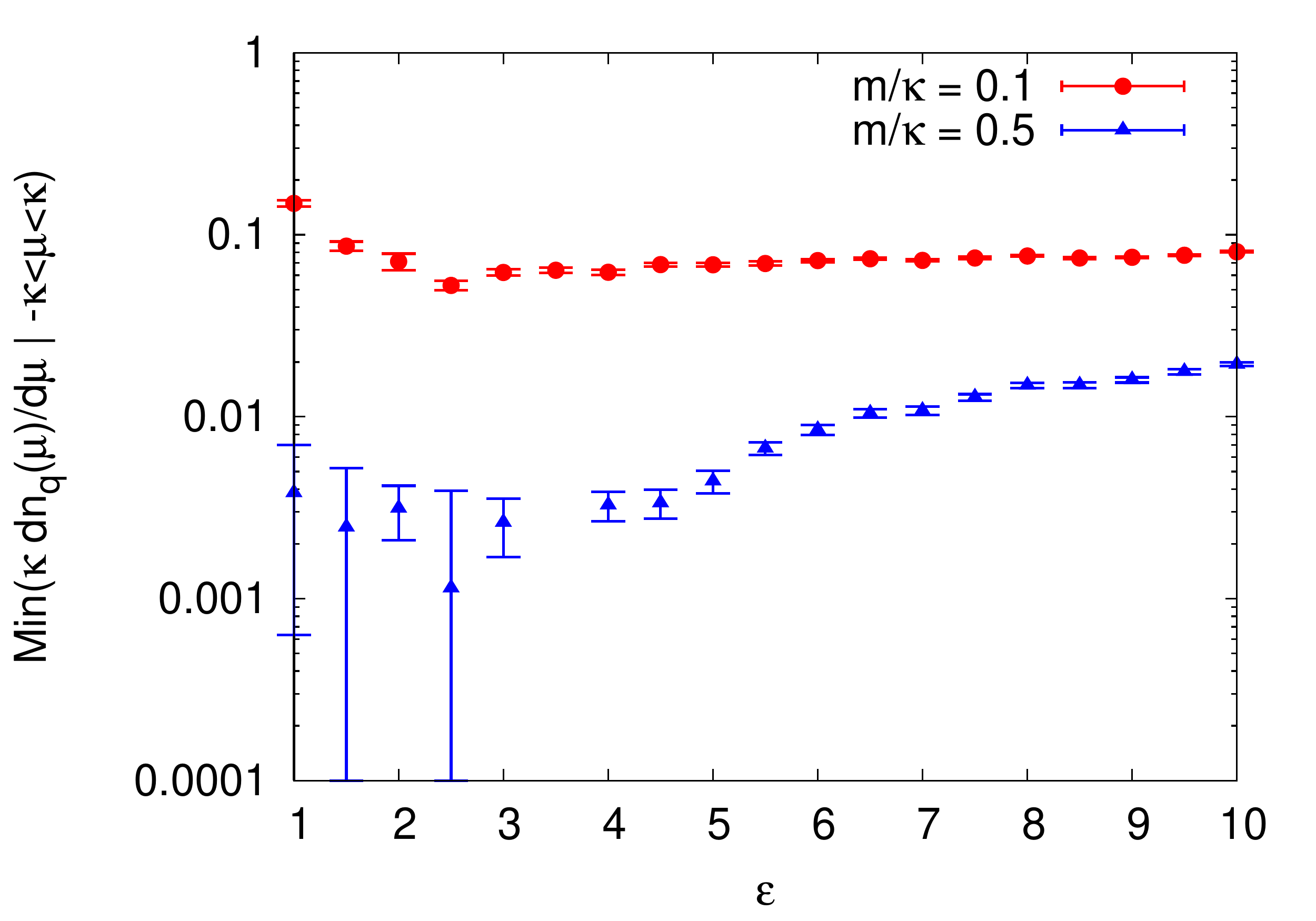}\\
  \caption{Minimal value of the estimate $\frac{d n_q\lr{\mu}}{d \mu}$ of the density of states for $\mu$ in the range $-\kappa < \mu < \kappa$ for different values of the staggered potential.}
  \label{fig:mindens}
\end{figure}

 Another important prediction of the renormalization-group analysis, which has been recently confirmed experimentally \cite{Elias:11:1}, is the logarithmic divergence of the Fermi velocity in the interacting tight-binding model \cite{Shankar:94:1}. As follows from (\ref{dos_dirac}), the increase of Fermi velocity should result in a depletion of the density of states near $E = 0$. Superficially, since the slope of $d n_q\lr{\mu}/d \mu$ near $\mu = 0$ clearly increases towards smaller $\epsilon$, it seems that our results point to the decrease of the Fermi velocity with interaction strength. However it might well be that due to the smearing (\ref{dndmu_vs_mu}) this increase of the density of states is just caused by the strong growth of the density of states at the Van Hove singularity. In order to demonstrate that the density of states near $E = 0$ indeed decreases, on Fig. \ref{fig:mindens} the minimal value of $d n_q\lr{\mu}/d \mu$ for $\mu$ in the range $-\kappa < \mu < \kappa$ is plotted as a function of $\epsilon$. This minimum is always situated close to $E = 0$. One can see that indeed this minimal value decreases as the Coulomb interaction becomes stronger for $\epsilon \gtrsim 4$. This is an indication of the decrease of the density of states in the vicinity of $E = 0$. As demonstrated in \cite{Lahde:09:2, Buividovich:12:1}, at smaller $\epsilon$ the symmetry between simple sublattices of graphene hexagonal lattice breaks down spontaneously. In this phase there is no reason to believe that the Fermi-Landau liquid model is a good approximation. A glance at Fig. \ref{fig:dndmu_k0.2000_s18_t18} suggest also that the width of the energy gap which is artificially induced by the staggered potential $m$ is practically independent of the interaction strength.

 To conclude, the presented results are consistent both with the sharpening of the Van Hove singularity and with the increase of the Fermi velocity predicted by renormalization-group calculations \cite{Shankar:94:1, Vozmediano:97:1}. It seems, however, that the former modification the dispersion relation is much stronger than the latter, and that much better energy resolution is required in order to study quantitatively the evolution of the Fermi velocity and the mass gap in the tight-binding model with interactions. At present larger-scale simulations with significantly smaller temperatures (so that the smearing factor in (\ref{dndmu_vs_mu}) is much narrower) and with larger lattices are in progress, which would hopefully help to study the quasiparticle dispersion relation in more details.

\section*{Acknowledgements}
 The author is grateful to Dr.~M.~I.~Polikarpov, Dr.~C.~Popovici and Dr.~M.~V.~Ulybyshev for many interesting discussions and also to Dr. L. von Smekal for his enlightening comments on the Van Hove singularity and the Lifshitz phase transition. Numerical calculations were performed at the ITEP computer systems ``Graphyn'' and ``Stakan'' (the author is grateful to A.~V.~Barylov, A.~A.~Golubev, V.~A.~Kolosov, I.~E.~Korolko, M.~M.~Sokolov for the help). The author was supported by the S. Kowalewskaja award from the Alexander von Humboldt foundation.


\end{document}